\documentclass[a4paper,fleqn]{cas-sc}
\usepackage{lineno,hyperref}
\usepackage{amsmath,amssymb}
\usepackage{url}
\usepackage{subcaption}
\usepackage{graphicx}
\usepackage{array,multirow} 
\usepackage{caption}
\modulolinenumbers[5]
\usepackage{array}
\usepackage{textcomp}
\usepackage{booktabs}
\usepackage{gensymb}
\usepackage{multirow}
\usepackage{float}
\usepackage{setspace} 

\usepackage[english]{babel}
\usepackage{blindtext}

\newcommand{\absdiv}[1]{%
  \par\addvspace{.5\baselineskip}
  \noindent\textbf{#1}\quad\ignorespaces
}

\usepackage[numbers]{natbib}
\newcolumntype{P}[1]{>{\centering\arraybackslash}p{#1}}

\begin{document}
\let\WriteBookmarks\relax
\def\floatpagepagefraction{1}
\def\textpagefraction{.001}

\shorttitle{Unet Architectures in Multiplanar Segmentation}    

\shortauthors{S.S. Sengar, C. Meulengracht, M.P. Boesen, A.F. Overgaard, H. Gudbergsen, J.D. Nybing, E.B. Dam}  

\title [mode = title]{Unet Architectures in Multiplanar Volumetric Segmentation - Validated on 3 Knee MRI Cohorts}  



\author[1]{Sandeep Singh Sengar}[orcid=0000-0003-2171-9332]



\ead{sengar@di.ku.dk}
\ead{+45-71645930}
\ead[url]{https://di.ku.dk/english/staff/?pure=en/persons/680742}


\affiliation[1]{organization={Department of Computer Science, University of Copenhagen},
            city={Copenhagen},
            postcode={2100}, 
            country={Denmark}}
\author[2]{ Christopher Meulengracht}
\author[2]{ Mikael Ploug Boesen}
\author[2]{ Anders Føhrby Overgaard}
\author[2]{ Henrik Gudbergsen}
\author[2]{ Janus Damm Nybing}
\author[1]{ Erik Bjørnager Dam}




\affiliation[2]{organization={Copenhagen University Hospital, Bispebjerg and Frederiksberg},
            city={Copenhagen},
            postcode={2400}, 
            country={Denmark}}

\cortext[1]{Corresponding author}



\begin{abstract}
\absdiv{\textbf{Background  and  Objective: }}The Unet has become the golden standard method for segmentation of 2D medical images that any new method must be validated against. However, in recent years, a number of variations to the seminal Unet has been proposed with promising results in the papers introducing them. However, there is no clear consensus if any of these architectures generalize as well and the Unet currently remains the methodological golden standard. The purpose of this study was to evaluate some of the most promising Unet-inspired architectures for the task of 3D segmentation.
\absdiv{Methods:}For segmentation of 3D scans, Unet-inspired methods are also dominant, but there is a larger variety across applications. By evaluating the architectures in a different dimensionality, embedded in a different method, and for a different task, we aimed to evaluate if any of these Unet-alternatives are promising as a new golden standard that generalizes even better than the Unet. Specifically, we investigated the architectures as the central 2D segmentation core in the Multi-Planar Unet 3D segmentation method that previously demonstrated excellent generalization in the MICCAI Segmentation Decathlon.
\absdiv{Results:}It would strongly support a claim of generalizability, if a promising Unet-variant consistently outperforms the Unet in this quite different setting. For this purpose, we evaluated four architectures for segmentation of cartilage from three different cohorts with knee MRIs. The implementation of our work is available at \href{https://github.com/sandeepsinghsengar/MPUNet2Plus}{link}.
\absdiv{Conclusions:}The proposed results support that the Unet2+ should in general be considered as another option to the highly popular method Unet.
\end{abstract}

\maketitle
\doublespacing

\section{Introduction}\label{sec:intro}
The encoder-decoder based UNet architecture introduced in 2015~\cite{ronneberger2015u} is highly popular for medical image segmentation as demonstrated by the one-citation-per-hour rate during 2020. The merits of the Unet include efficient information flow through skip connections between encoder and decoder, good performance with limited data, and impressive generalization across diverse tasks~\cite{drozdzal2016importance, zhuang2018laddernet}. In recent years, several variations of UNet architectures have been proposed with  promising results for 2D images~\cite{yahyatabar2020dense, mahmud2021polypsegnet}. Among them, UNet2+, UNet3+, and deep supervised versions of UNet2+ (UNet2+ DS) and UNet3+ (UNet3+ DS) are the most popular UNet inspired architectures for medical imaging~\cite{zhou2018unet++, huang2020unet}. UNet 2+~\cite{zhou2018unet++} and UNet 3+~\cite{huang2020unet} have been developed to reduce the fusion of semantically non-similar features from plain skip connections in UNet. These architectures employ the nested and dense skip connections to strengthen the connections and to decrease the semantic gap between encoder and decoder. Furthermore, the results can be improved by extracting features from multiple deep layers. Numerous skip connections based approaches have been developed for aggregating and integrating multi-level features. Moreover, deep supervision (DS) is also employed in UNet2+ and UNet3+ models to learn hierarchical representations from the full-scale aggregated feature maps. The performance of both deep supervised and non-deep supervised versions of UNet 2+ and UNet 3+ have been demonstrated to be superior to the original UNet for certain 2D medical image datasets~\citep{huang2020unet}. In-spite the success of these models, it is not clear if these architectures generalize well to other tasks.

As evident from numerous segmentation algorithms~\cite{chen2017deeplab}, \cite{zhou2019unet++}, \cite{feng2020cpfnet} distinctive information can be explored through feature maps in different scales. The position of organs can be detected through high-level semantic features, while the boundaries of organs can be obtained through low level detailed features. In addition to these, rich features can be captured through different orientations of 3D MRI image. Considering these points, we have selected the 3D image segmentation method MPUNet~\cite{perslev2019one} as a baseline model for the further verification of different UNet variants. MPUNet secured rank $5^{th}$ and $6^{th}$ in the first and second round of the 2018 Medical Segmentation Decathlon. The general architecture is the reason for selection of this method, i.e. the system requires no task-specific information, no human interaction, and is based on a fixed model topology and a fixed hyperparameter set. We have applied four different architectures -- UNet2+, UNet3+, UNet2+ DS, and UNet3+ DS -- in the MPUNet method and called these MPUNet2+, MPUNet3+, MPUNet2+DS, and MPUNET3+DS respectively.  

Cartilage segmentation from knee MRIs is an important step in clinical research of the progression of osteoarthritis and likely in future clinical practice. We evaluated the performance of the proposed architectures on three knee MRI cohorts from the Osteoarthritis Initiative (OAI), the Center for Clinical and Basic Research (CCBR), and the Liraglutide trial (LIRA). Using three cohorts with very different MRI sequences provides a challenging evaluation.

The rest of this paper is organized as follows. After this introductory section, brief introduction of related works is given in Section~\ref{sec:related}. The proposed method is elaborated in Section~\ref{sec:method}. Experimental setup and results are provided in Section~\ref{sec:data}. The important discussion related to paper is given in Section~\ref{sec:discusion}. Finally, conclusion of work is provided in Section~\ref{sec:Conclusions}.

\section{Related works}\label{sec:related}
Knee osteoarthritis is a whole joint disease~\cite{loeser2012osteoarthritis} originated by a multifactorial combination of biochemical~\cite{sokolove2013role}, biomechanical~\cite{englund2010role}, systemic and intrinsic~\cite{warner2016genetics} risk factors. Ambellan et al.~\cite{ambellan2019automated} enhanced the Liu et al.~\cite{liu2018deep} approach by adapting the idea of slice-wise segmentation and adding Statistical Shape Model (SSM) as extra feature into their 2D/3D U-net based bone segmentation technique. Here, SSM is employed to (1) remove false positive voxels from tibia and femur (2) fill holes in segmentation masks due to poor intensity contrast. However, the good performance of this model is achieved through huge computational power and localized training. As per Ambellan et al.~\cite{ambellan2019automated}, the total duration of deep learning model implementation is 43 weeks, which shows the expansive cost of computation. In literature, many of the researcher simplified CNN architecture to reduce complexity, but the issue is still exist. The segmented cartilages from Knee MRI images produces are employed in a broad range of OA-related studies for example (1) classification and detection of knee OA progression (Ashinsky et al.~\cite{ashinsky2015machine}; Tiulpin et  al.~\cite{tiulpin2019multimodal}; Ashinsky et al.~\cite{ashinsky2017predicting}; Chang et al.~\cite{leung2020prediction}; Almajalid et  al.~\cite{almajalid2019identification}), imaging biomarkers analysis (Hafezi-Nejad et al.~\cite{hafezi2017prediction}; Schaefer et al.~\cite{schaefer2017quantitative}; Shah et al.~\cite{shah2019variation}; Williams et al.~\cite{williams2010automatic}), biomechanical modeling (Liukkonen et  al.~\cite{liukkonen2017application}) and stimulation of cartilage degeneration (Peuna et al.~\cite{peuna2018variable}; Liukkonen et al.~\cite{liukkonen2017simulation}; Mononen et al.~\cite{mononen2019utilizing}). Cartilage segmentation is still an active research problem from long time due to thin (in few millimetres) and extremely thin (in submillimetre) structure of knee cartilages. Furthermore, patella, femoral, and tibial cartilages have distinctively different and changing shapes across the slides. Perslev~\cite{perslev2019one} proposed a simple and thoroughly evaluated multiplanar UNet (MPUNet) for segmentation of arbitrary medical image volumes. This model requires no human interaction, no task-specific information, is based on a fixed hyperparameter set, a fixed model topology and, eliminating the process of model selection and its inherent tendency to cause method-level overfitting. Furthermore, Xu~\cite{xusegmentation} employed the MPUNet work to segment the 3D jaw images. Deep multi-planar co-training based semi-supervised 3D abdominal multi-organ segmentation is proposed by Zhou et al.~\cite{zhou2019semi}. The relevance of multi-planar MRI acquisitions is assessed for prostate segmentation using deep learning techniques~\cite{lozoya2018assessing}.  This method is depended on an ensemble of convolutional neural networks, each independently trained on a single imaging view.

\section{Method} \label{sec:method}

The multiplanar Unet (MPUNet)~\cite{perslev2019one} works by traversing the 3D volume from different orientations. Thereby, multiple 2D image planes are visited. During training, a standard UNet is trained on these 2D images. During inference, the 3D volume is traversed resulting in a prediction volume from each traversal orientation. These prediction volumes are linearly fused into a single segmentation volume.  The steps are explained in a bit more detail below. 

\subsection{Pre processing} \label{processing}

There is a minimum of image preprocessing applied outside of the network architectures. An image and channel wise outlier robust preprocessing has been applied to scales the intensity values. This preprocessing is done based on inter-quartile and median range calculated over foreground voxels. Here, foreground voxels are separated by having intensities greater than the first percentile of the intensity distribution. 

\subsection{Input data generation} 

\begin{figure}[ht]
  \centering
    \includegraphics[height=6cm, width=12cm]{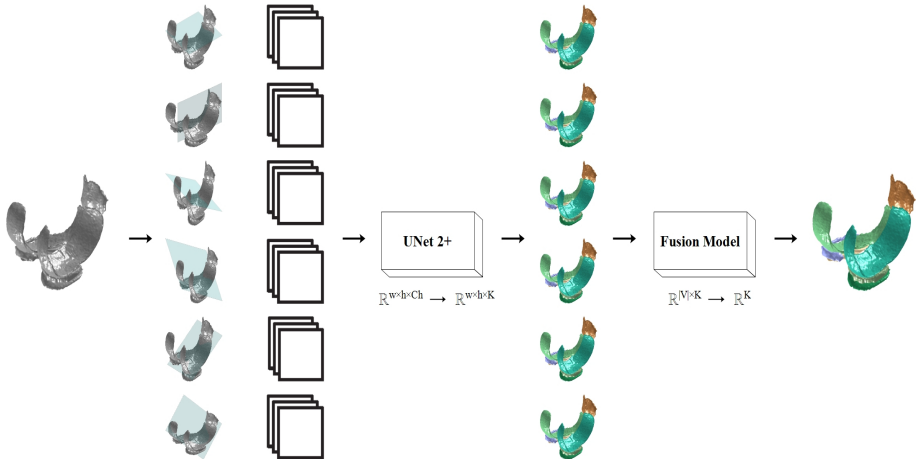}
      \caption{Block diagram of the MPUNet2+ Model.}
      \label{block}
\end{figure}

The first target of multiplanar method is to generate the input 2D slices from 3D volume for training and testing steps~\cite{perslev2019one}. These slices have been generated by sampling from multiple planes of random orientation spanning the 3D image. The advantage of data collected from multiple planes is that model will learn to segment the input captured from diverse views, see Fig.\ref{block}. To generate the 2D inputs for the model, we define a set $V = \{v_1, v_2,...,v_k\}$ of $k$ randomly sampled unit vectors in $R^3$. The axes through 3D image along which we sample 2D slices are defined by this set $V$ i.e. we can get k sets of 2D slices $L^V=\{(x_n^V,y_n^V)\}_{n=1}^{N_V}$, where $N_V$ is the number of 2D slices obtained from plane $V$; $x$ and $y$ are the 2D slice and it's corresponding label respectively; $y_i^V \in \{0, 1, . . . , K\}$ is the cartilage label (0 means background) of the $i^{th}$ pixel in $x$. In our case, we have selected the value of $k \in \{1, 3, 6\}$ i.e. suppose $k=6$ then we re-sampled the slices from 6 different random angles, similarly for 1 plane and 3 planes. Here, for each pair of random vectors, there are at least $60^{\circ}$  difference between them. For one plane, we have considered the sagittal axis, because most of the relevant features can be obtained through it. Due to lack of GPU computation, we have used only $k = 3, 6$ for the reported evaluations of MPUNet2+ deep, MPUNet3+ deep models, and all models with LIRA dataset. These planes have been selected based on experiments over OAI and CCBR datasets using MPUNet, MPUNet2+, and MPUNet3+ models i.e. best performer among these.

\subsection{Model} \label{model}

As shown in Fig.~\ref{block}, model $f(x; \theta)$ accepts  as input multi-channel 2D image slices of size $w\times h$,
$x \in \mathbb{R}^{w \times h \times Ch}$ and generates  a probabilistic segmentation map $S \in \mathbb{R}^{w \times h \times K}$ for
K classes. At the heart of our system lies UNet2+, UNet3+, UNet2+ DS, and UNet3+ DS; these are powerful architectures motivated by UNet~\citep{zhou2018unet++,huang2020unet}. As comparison to other variants of UNet architectures, UNet2+~\cite{zhou2018unet++} combines the multi-scale features by employing a dense network of re-designed skip connections as an intermediary grid between the encoder and decoder. This full-scale skip connections architectures establish the interconnection between the encoder and decoder as well as intraconnection among the decoder sub-networks. This arrangement of full-scale skip connections minimizes the loss of semantic information between the encoder and decoder. UNet3+~\cite{huang2020unet} is also a another variants of UNet with lots of skip connections, but lesser than UNet2+. UNet2+ DS~\citep{zhou2018unet++} and UNet3+ DS~\cite{huang2020unet} add full-scale deep supervision in the obtained aggregated feature maps of UNet2+ and UNet3+ respectively to learn hierarchical representations.

\begin{figure}[]
	\begin{tabular}{ccc}
		\begin{subfigure}{0.33\textwidth}\centering\includegraphics[height=4cm, width=4cm]{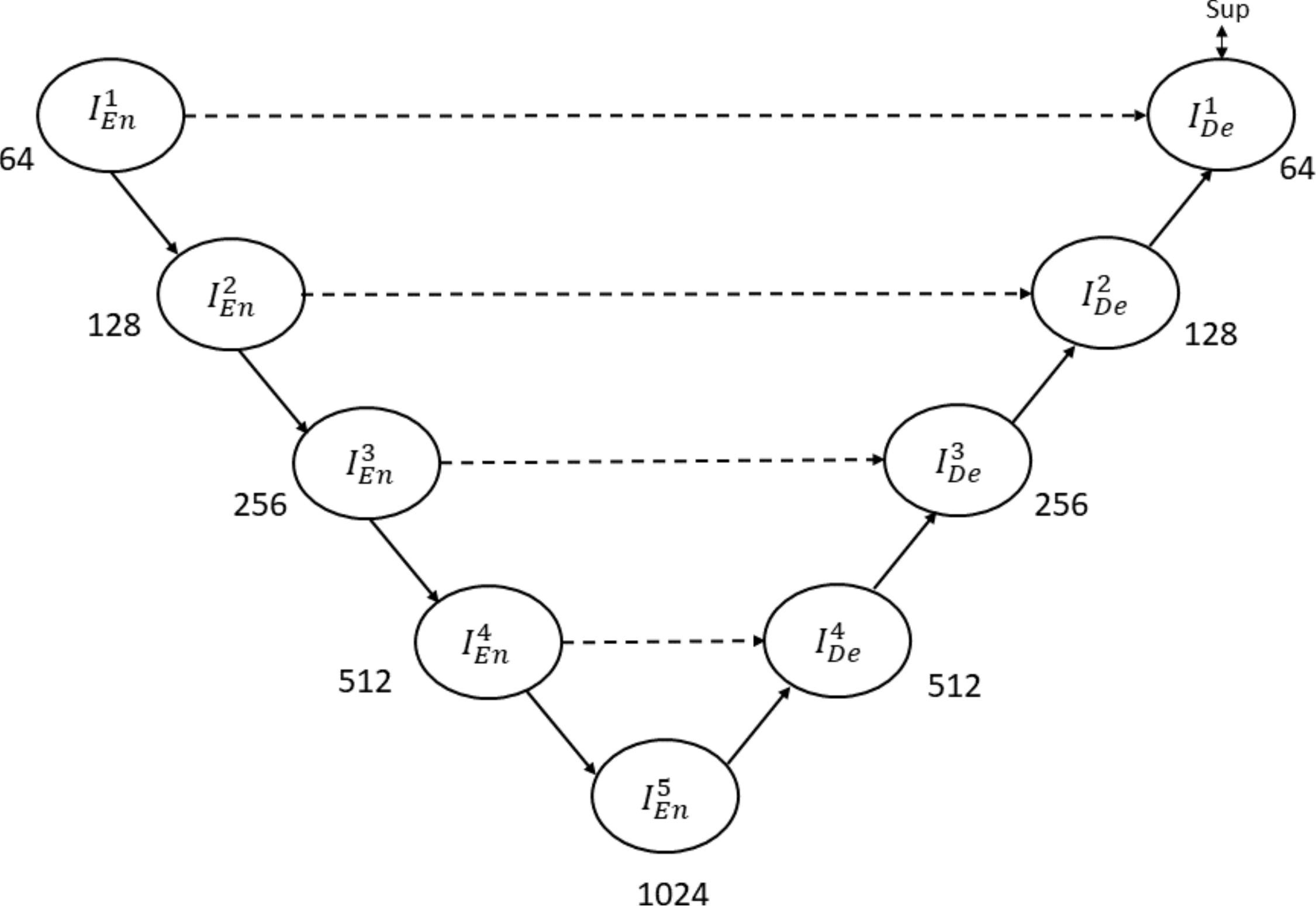}\caption{}\end{subfigure}& 
		\begin{subfigure}{0.33\textwidth}\centering\includegraphics[height=4cm, width=4cm]{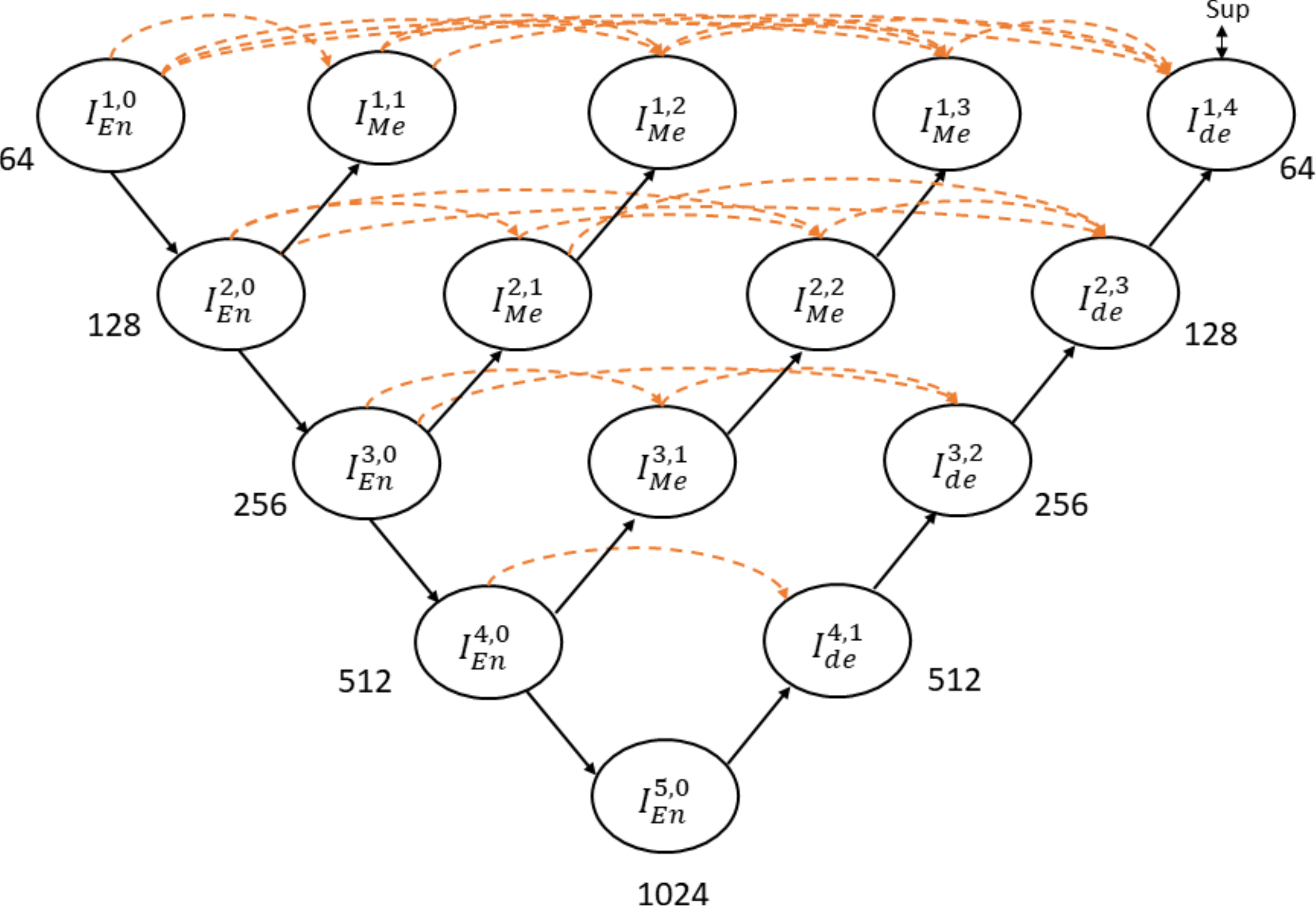}\caption{}\end{subfigure}&
		\begin{subfigure}{0.33\textwidth}\centering\includegraphics[height=4cm, width=4cm]{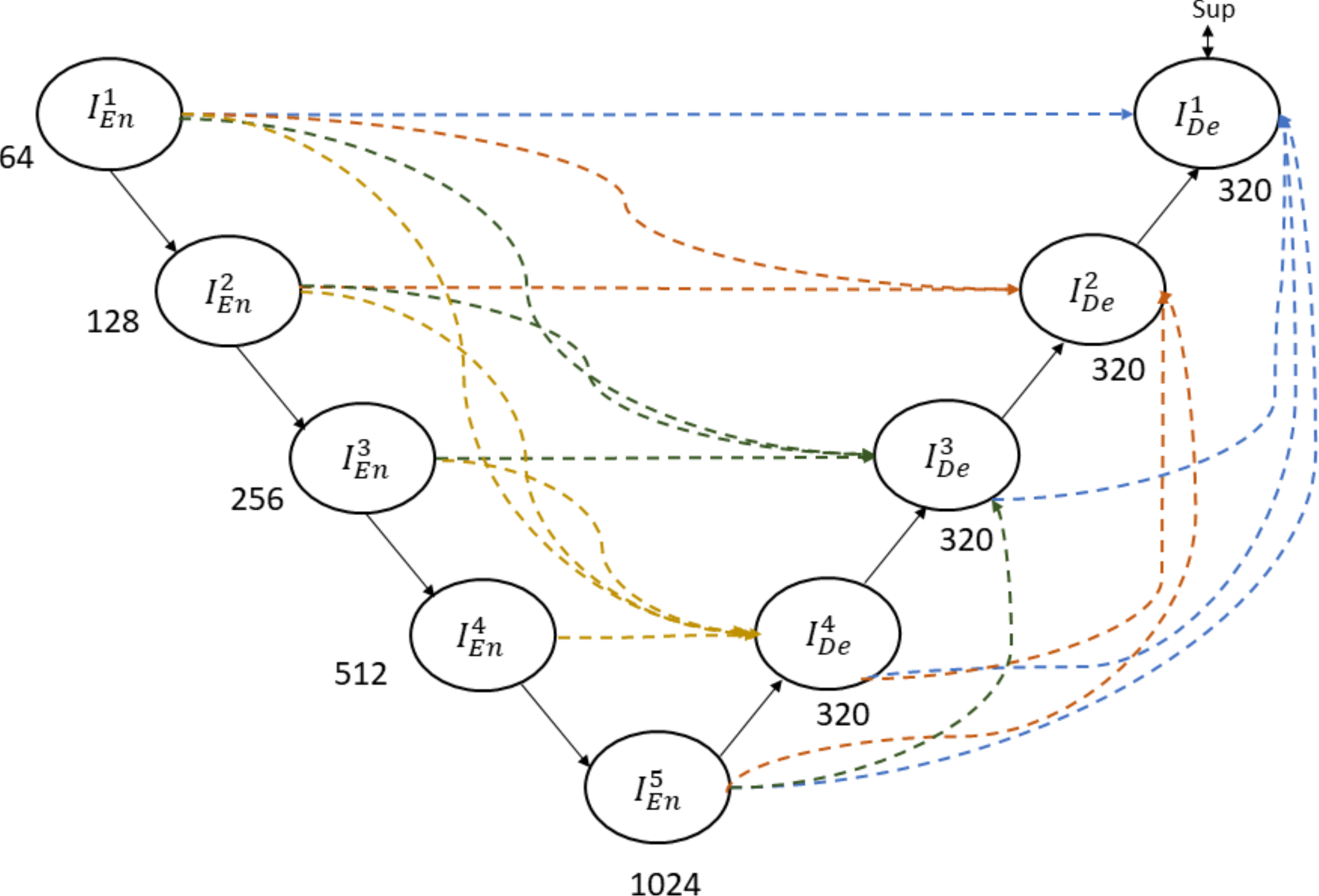}\caption{}\end{subfigure}
	\end{tabular}
	\caption{ \label{unet_models}
		Architecture of (a) UNet; (b) UNet2+; (c) UNet3+. The depth of each node is displayed under the circle.} 
\end{figure}

Perslev et al.~\citep{perslev2019one} have increased the number of filters by a factor of $\sqrt{2}$ in their UNet architecture, due to that MPUnet has higher number of parameters, which leads to high computational cost. The model architecture of UNet, UNet2+, and UNet3+ can be seen in Figs.~\ref{unet_models}(a), \ref{unet_models}(b), and \ref{unet_models}(c) respectively. For the comparison purpose, we have adopted deep supervision version of UNet2+ and UNet3+ from Huang et al.~\cite{huang2020unet} and architecture of these models can be referred from here.\\
It is good to mention here that employed UNet2+ and UNet3+ are more efficient with fewer parameters. Encoder part of Unet2+ and UNet3+ architectures share the same structure like UNet, where $I_{En}^{j}$ has $32\times 2^{i}$ channels.
The number of parameters in $j^{th}$ decoder stage of Unet2+ can be calculated as:
\begin{equation}
\label{unet++}
    P_{U2+}^j=F\times F \times \Big[d(I_{De}^{j+1}) \times d(I_{De}^{j})+d \Big(I_{En}^j +\sum_{m=1}^{N-1-j} I_{Me}^{j,m}+ I_{De}^j\Big)\times d(I_{De}^{j})\Big] 
\end{equation}
Where $F$, $d(\cdot)$, and N are the convolution kernel size (length), depth of the nodes, and total number of levels respectively. \\
Similarly, the number of parameters in $j^{th}$ decoder stage of UNet3+ can be calculated as:
\begin{equation}\label{unet3+}
    P_{U3+}^j=F\times F \times \Big[ \Big( \sum_{m=1}^{j} d(I_{En}^m)+ \sum_{m=j+1}^{N} d(I_{De}^m)\Big) \times 64 +d(I_{De}^{j})^2 \Big]
\end{equation}
For the comparison purpose, the number of parameters in $j^{th}$ decoder stage of UNet can be calculated as:
\begin{equation}
\label{unet}
    P_{U}^j=F\times F \times \Big[d(I_{De}^{j})\times d(I_{De}^{j+1})+ d(I_{De}^{j})^2+d(I_{En}^j + I_{De}^j) \times d(I_{De}^j) \Big]
\end{equation}

\subsection{Training, fusion and testing} \label{training}

For the training, batches of images are provided to the model. These batches are formed from the $k$ planes in set $V$ without giving information about the corresponding axis. During inference, the model predicts along each plane generating a set of $k$ segmentation volumes $S=\{S_v \in \mathbb{R}^{w\times h \times d \times K} | v\in V\}$. To get point correspondence, every $S_v$ is mapped to the input space through allocating to each voxel in the input data the value of its closest predicted point in $S_v$. Distances are calculated in physical coordinates. \\

At the test time, linear fusion model is applied to combine the numerous candidate segmentations for each subject. These numerous candidate segmentations are obtained through segmenting the entire 3D data from each plane, see Fig.~\ref{block}.  A weighted sum of the per-plane and per-class softmax scores are employed to map S to a single probabilistic segmentation. For all $w \cdot h\cdot d$ voxels x in S and each class $c \in \{1,...,K\}$ the fusion model $f_{fusion} :\mathbb{R}^{|V|\times K} \rightarrow \mathbb{R}^K$ computes $z(x)_c=\sum_{k=1}^{|V|}W_{k,c}\cdot p_{k,x,c}+\beta_c$. Where, $p_{k,x,c}$ represents the probability of class $c$ at voxel x as predicted by segmentation $S_k$. The $\beta \in \mathbb{R}^K$ is bias parameter, which is used to correct the whole tendency to predict a particular class and $W \in \mathbb{R} ^{|V| \times K}$ employed to weighs the probabilities of individual class as predicted from each plane. The validation data is used to learn the $f_{fusion}$ parameters. When volume will be seen from different directions (in our case $|V|$ directions), different classes may seem in different levels of recognizability and shapes. Due to aforementioned cause, the model scales the prediction based on the views performance on each class.

The effect of applying the affine transformations on the 3D volume and giving the transformed data to a model is similar to processing the image data from different views. Therefore, all the models in our method are with systematic affine data augmentation. Non-linear transformations are applied along with multi-view sampling to augment the training data. Each sampled image in a batch is augmented by the Random Elastic Deformations algorithm~\cite{simard2003best} with a probability of 1/3.  To produces the augmented images with large variability in respect of smoothness and deformation strength, the deformation intensity multipliers ($\sigma$) and elastic constants ($\alpha$) are sampled uniformly from [0, 450] and [20, 30] respectively. 
\subsection{Datasets}

For the experimentation purpose, we have used Knee MRI images with independent manual segmentations acquired from the Osteoarthritis Initiative (OAI)~\cite{OAI}, CCBR (the Center for Clinical and Basic Research)~\cite{dam2015automatic}, and LIRA~\cite{gudbergsen2021liraglutide}. All three cohorts were randomly divided with the ratio of 60:20:20 for training, testing and validation in a 5 fold cross validation. 

\noindent \textbf{OAI:} OAI is a multi-center, ten-year observational study of men and women of age between 45-79 years and baseline body mass index (BMI) $26.9 \pm 4.44$. It is sponsored by the National Institutes of Health. The targets of this organization are to deliver resources to empower a better understanding of prevention and treatment of knee osteoarthritis, one of the common reasons of disability in adults. We have employed 60 Knee MRI volumes from OAI. 

\noindent \textbf{CCBR:} This dataset is a community study of men and women of age between $54\pm 15$ and BMI $26 \pm 4$ from the greater Copenhagen region. The cohort included healthy and subjects with mild to advanced radiographic OA. We have included 140 Knee MRI volumes from CCBR, which are randomly divided among training, testing and validation for cross validation. 

\noindent \textbf{LIRA:} This is a sub-study to a randomised trial (LOSEIT) investigating the effect of liraglutide on body weight and pain in overweight or obese patients with knee osteoarthritis (ClinicalTrials.gov NCT02905864). Eligibility criteria for the patient trial were as follows: age: 18 - 74 years, BMI $\ge$ 27 kg/m2; clinical diagnosis of KOA, confirmed by radiology but restricted to definite radiographic OA at early to moderate-stages (Kellgren-Lawrence grades 1, 2, or 3), motivation for weight loss. All patients were subjected to an 8-week diet intervention with a minimum 5\% initial weightloss, after which the patients were randomised to receive either 3 mg liraglutide or 3 mg placebo for an additional 52 weeks. MRI was performed at 8 weeks and at randomization week 0 and again after 52 weeks. We have used 39 knee MRI scans from LIRA.\\

\noindent The Description of these datasets are shown in Table~\ref{datset}.

\begin{table}[]
\begin{center}
\caption{Description of OAI, CCBR, and LIRA datasets.}
\resizebox{\columnwidth}{!}{
\begin{tabular}{|l|l|l|l|}
\hline
\textbf{Dataset}             & OAI      & CCBR        & LIRA \\ \hline
\textbf{No. of scans}     & 60      & 140          &  39    \\ \hline
\textbf{No. of slices}     & 9600      & 14560          &  6240    \\ \hline
\multicolumn{1}{|l|}{\begin{tabular}[c]{@{}l@{}} \textbf{Cartilage} \\\textbf{information}\end{tabular}} &
  \multicolumn{1}{l|}{\begin{tabular}[c]{@{}l@{}}Medial Tibial, Lateral Tibial, \\ Medial Femoral, Lateral \\ Femoral,  Patellar, Lateral \\ Meniscus, Medial Meniscus\end{tabular}} &
  \multicolumn{1}{l|}{\begin{tabular}[c]{@{}l@{}}Medial \\ Tibial, \\ Medial \\ Femoral\end{tabular}} &
  \multicolumn{1}{l|}{\begin{tabular}[c]{@{}l@{}}Lateral Femoral, \\ Medial Femoral, \\ Tibia, Lateral Tibial, \\ Medial Tibial\end{tabular}} \\ \hline
\textbf{Scannner}            & Siemens 3T     & Esaote 0.18T &  Siemens 3T \\ \hline
\textbf{Protocol}            & DESS     & Turbo 3D T1 &  DESS    \\ \hline
\textbf{FOV}                 & 140 mm   & 180 mm       &   140 mm \\ \hline
\textbf{Slice thickness/gap} & 0.7/0 mm & 0.8/0 mm     &  0.7/0 mm    \\ \hline
\textbf{Flip angle}          & 25\textdegree       & 40\textdegree           &   25\textdegree   \\ \hline
\textbf{X-resolution}        & 0.365 mm & 0.7 mm       &0.365 mm      \\ \hline
\textbf{Y-resolution}        & 0.365 mm & 0.7 mm       & 0.365 mm     \\ \hline
\textbf{Slice size}          & 384$\times$ 384  & 256$\times$ 256      &  384$\times$ 384    \\ \hline
\end{tabular}}
\label{datset}
\end{center}
\end{table}

\section{Experimental setup and results} \label{sec:data}

\subsection{Experimental setup}
For better generalization, different multiplanar based UNet architectures have been compared. The level of all the tested models are kept five. To do the better comparison, all experiments have been performed using 5-fold cross-validation with multiple planes (1, 3, and 6). All models have been implemented in TensorFlow 2.3.0, trained on a Nividia-Titan X graphics processing unit with 8GB memory using Adam optimizer~\citep{zhang2018improved} with a batch size of 4-16 (16 by default, it has been reduced by 2 until batches fit in GPU memory), batch normalization~\citep{ioffe2015batch} between successive layers, for a maximum of 500 epochs and a patience of 15 epochs based on the validation accuracy. Model with the maximum dice score is employed for testing. The learning rate for each models has been kept $5\times10e-5$ to overcome exploding gradient problem at larger learning rates. The values of other Adam optimizer parameters $\beta_1$, $\beta_2$ and $\epsilon$ are fixed 0.9, 0.99, and $1\times 10e-8$ respectively. The probability to perform data augmentation is 0.33. We have used the cross entropy loss function and $L_2$ regularization for the implementation purpose. The number of training and validation images per epoch are 2500 and 3500 respectively. All the non-constant hyper-parameters can automatically be deduced from the training data. There is no post-processing step in our method, because it is generally task specific. Our implementation extends the publicly available MPUnet source code at~\href{https://github.com/perslev/MultiPlanarUNet}{link}. The source code for this study is all available as open source. See the GitHub repository \href{https://github.com/sandeepsinghsengar/MPUNet2Plus}{here} for our implementation details.

\subsection{Results}

We have done both quantitative and qualitative evaluations for the comparison purpose. 

\subsubsection{Quantitative evaluation}

\noindent The following steps have been followed to generate the dice score and computation time:\\
\noindent \textbf{Dice Score:} a) split the data-set into five randomly generated folds of training, testing and validation with ratio of 60:20:20 respectively (b) extract the 2D slices from the 3D data-set with different planes (1, 3, 6 number of planes). For the case of one plane, we are considering sagittal axis (c) apply 5-fold cross validation and generate result (dice Score). \\
\noindent \textbf{Computation time:} Computation time is the mean of all five folds corresponding to total training time. 

We have performed experiments using 1, 3, and 6 planes. Due to limited computing resources, we have experimented 1-plane only for MPUNet, MPUNet2+, and MPUNet3+ on OAI and CCBR datasets. The 1-plane dice scores for MPUNet, MPUNet2+, and MPUNet3+ are 0.847, 0.872, and 0.852 respectively for OAI and 0.795, 0.815, and 0.817 respectively for CCBR datasets. As these dice scores are not significantly better than corresponding 3-planes and 6-planes, therefore we neglected 1-planes for further experiments. The result of the proposed method along with compared techniques in terms of dice score, computation time, and number of parameters (\#param.) across all datasets (for 3-planes and 6-planes) are reported in Table~\ref{performance}. It is evident from Table~\ref{performance}, MPUNet2+ considerably perform better among all. Further to verify the significant performance of MPUNet2+, the obtained dice scores from 6-planes are validated statistically using the paired t test and the Wilcoxon rank-sum test at 5\% level of significance for OAI, CCBR and LIRA datasets. The null hypothesis of the test assumes that there is no significant difference between the two sets of results obtained by the proposed method and the other compared methods. Paired t-test and Wilcoxon rank-sum test results obtained by the MPUNet2+ versus other multiplanar approaches in terms of p-score are reported in Tables~\ref{ttest} and~\ref{wtest} respectively. Results of these tests are statistically significant (at 5\% level of significance) if the corresponding \textit{p} value is less than or equal to 0.05, indicating that the null hypothesis is rejected, that means there is statistically significant difference in the results obtained by the compared and comparing algorithms. Results of the paired t-test and Wilcoxon rank-sum test suggest that the dice score obtained by the proposed MPUNet2+ approach significantly dominates the other counterpart methods, except in case of MPUnet and MPUNet3+ for OAI and CCBR datasets respectively (for paired t-test) and MPUNet3+ for CCBR dataset (for Wilcoxon rank-sum test)[in Table~\ref{performance_all}, significantly better and near to significantly better results are highlighted in bold].

\begin{figure}[ht]
  \centering
    \includegraphics[width=.7\linewidth]{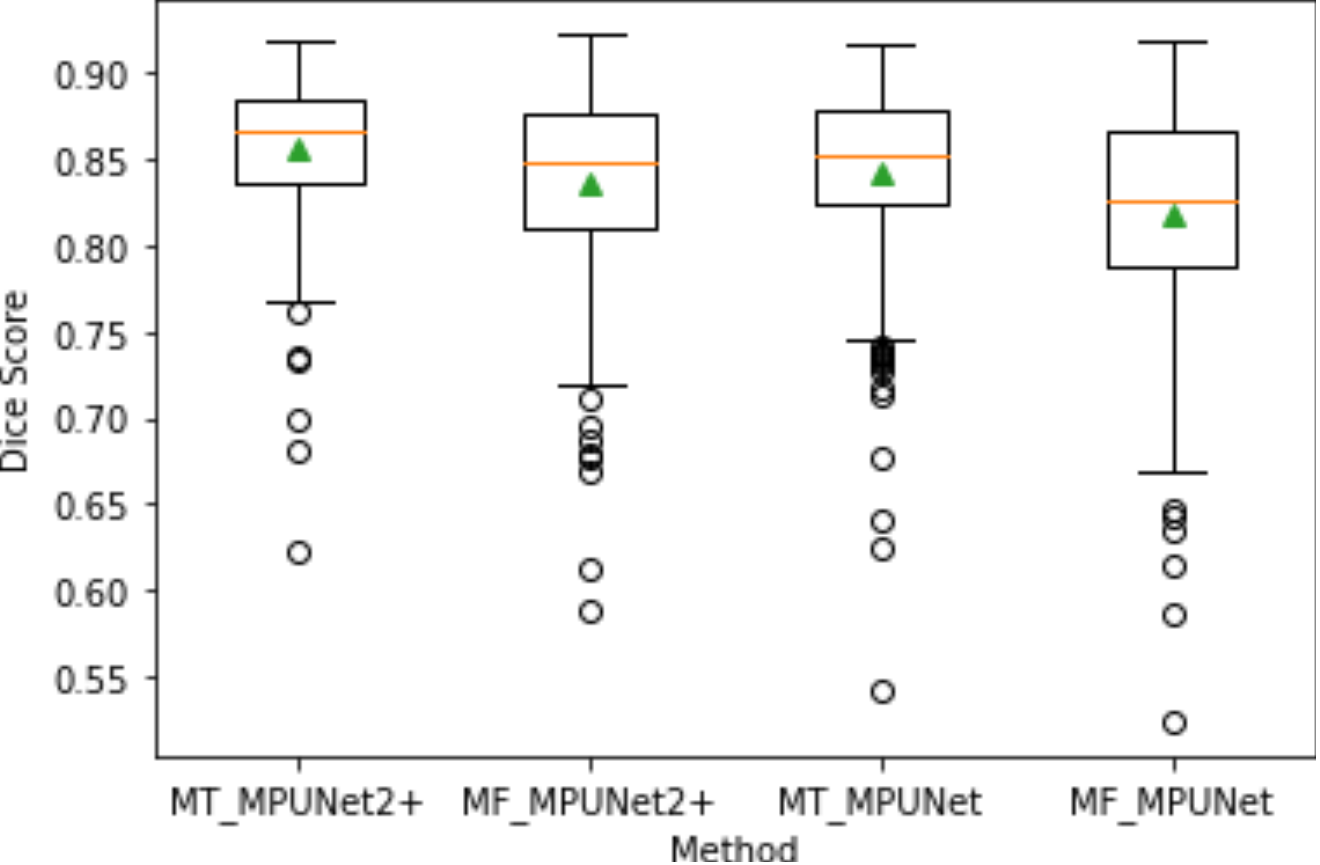}
      \caption{Box and whisker plot of the dice score on medial tibial and medial femoral cartilages of combined OAI, CCBR, and LIRA datasets for MPUNet and MPUNet2+ methods with 6 planes}
      \label{box}
\end{figure}

It is evident from Table~\ref{performance_all} that, Computation time and number of parameters in MPUNet2+ are fewer than MPUNet. Moreover, the computation time and number of parameters are considerably higher in 6 planes version of MPUNet2+ than MPUNet3+, MPUNet2+ DS, and MPUNet3+ DS, but dice score of MPUNet2+ is far better than these approaches. It is proved from the aforementioned discussion that MPUNet2+ outperform the other models in terms of dice score, number of parameters and computational time.\\
To better generalization of the approach, We have employed box and whisker plot (Fig.~\ref{box}) to do cartilage-wise comparison between MPUNet and MPUNet2+. Here, we have combined the dice scores of all the three data-sets (OAI, CCBR, and LIRA) corresponding to medial tibial and medial femoral cartilages (as only these two cartilages are common in all the three data-sets). In Fig.~\ref{box}, green triangle represents the mean dice score corresponding to particular cartilage. The statistical description of Fig.~\ref{box} are shown in Table~\ref{Statistical_box}. It is evident from aforementioned table that minimum, maximum, mean, median, 25th and 75th percentile values are higher in MPUNet2+ as compared to MPUNet for both the cartilages. Furthermore, there are higher number of outliers in MPUNet than MPUNet2+. \\
In addition to above, we have displayed the cartilage-wise dice score of MPUNet2+ over the individual dataset in Table~\ref{Cartilage}. It is shown from table that, the individual cartilage also performs well. 

\begin{table}[]
\begin{center}
\caption{Performance comparison on OAI, CCBR, and LIRA datasets. The results have been computed with 3, and 6 number of planes. The performance is reported as mean dice score and training time (in hour) averaged over 5-folds. Here, displayed mean dice score is calculated over non-background classes only. Number of parameters ($\#$param.) for all methods are also reported. Highest DICE scores are highlighted in \textbf{bold} together with those that are not statistically significantly different.} \label{performance_all} 
\resizebox{\columnwidth}{!}{
\begin{tabular}{|l|c|c|c|c|c|}
\hline
\multirow{2}{*}{\textbf{Model}} & \multicolumn{2}{c|}{\textbf{3 Planes}} & \multicolumn{2}{c|}{\textbf{6 Planes}} & \multirow{2}{*}{\textbf{\#Param.}} \\ \cline{2-5}
            & \textbf{dice Score}                 & \textbf{time (h)} & \textbf{dice Score}                  & \textbf{time (h)} &          \\ \hline
\multicolumn{6}{|c|}{\textbf{Dataset: OAI, Number of classes: 7}}                                            \\ \hline
MPUNet      & .850           & 93       & \textbf{.868}  & 67       & $\approx62 M$ \\ \hline
MPUNet2+    & \textbf{.873 } & 26       & \textbf{.867}            & 32       & $\approx36 M$ \\ \hline
MPUNet3+    & .861  & 47       & \textbf{.869 }  & 35       & $\approx27 M$ \\ \hline
MPUNet2+ DS & .854           & 24       & .846            & 24       & $\approx36 M$ \\ \hline
MPUNet3+ DS & .854           & 57       & .856           & 55       & $\approx27 M$ \\ \hline
\multicolumn{6}{|c|}{\textbf{Dataset: CCBR, Number of classes: 2}}                                           \\ \hline
MPUNet      & .817           & 15       & .816            & 16       & $\approx62 M$ \\ \hline
MPUNet2+    & .830  & 18       & \textbf{.842 }  & 16       & $\approx36 M$ \\ \hline
MPUNet3+    & \textbf{.836 } & 18       & \textbf{.841  } & 20       & $\approx27 M$ \\ \hline
MPUNet2+ DS & .827           & 18       & .836            & 14       & $\approx36 M$ \\ \hline
MPUNet3+ DS & .824           & 17       & .832            & 19       & $\approx27 M$ \\ \hline
\multicolumn{6}{|c|}{\textbf{Dataset: LIRA, Number of classes: 5}}                                           \\ \hline
MPUNet      & .832          & 24       & .864  & 34       & $\approx62 M$ \\ \hline
MPUNet2+    & \textbf{.861 } & 26       & \textbf{.868 }  & 37       & $\approx36 M$ \\ \hline
MPUNet3+    & .845           & 29       & .850            & 26       & $\approx27 M$ \\ \hline
MPUNet2+ DS & .835           & 23       & .844            & 20       & $\approx36 M$ \\ \hline
MPUNet3+ DS & .853  & 34       & .856            & 43       & $\approx27 M$ \\ \hline
\end{tabular}}
\label{performance}
\end{center}
\end{table}

\begin{table}[]
\begin{center}
\caption{P-values from paired t-test performed on the Dice score obtained by the MPUNet2+ versus other compared methods (viz., MPUNet, MPUNet3+, MPUNet2+ DS, and MPUNet3+ DS at 6-planes) in terms of p-score for OAI, CCBR, and LIRA datasets.} 
\label{ttest} 
\resizebox{\columnwidth}{!}{
\begin{tabular}{|l|c|c|c|c|}
\hline
\textbf{Dataset} &
  \textbf{\begin{tabular}[c]{@{}l@{}}MPUNet2+ \\ vs.\\ MPUNet\end{tabular}} &
  \textbf{\begin{tabular}[c]{@{}l@{}}MPUNet2+ \\ vs.\\ MPUNet3+\end{tabular}} &
  \textbf{\begin{tabular}[c]{@{}l@{}}MPUNet2+ \\ vs.\\ MPUNet2+ DS\end{tabular}} &
  \textbf{\begin{tabular}[c]{@{}l@{}}MPUNet2+ \\ vs.\\ MPUNet3+ DS\end{tabular}} \\ \hline
OAI  & 0.187    & 0.045    & 1e-10 & 2e-18 \\ \hline
CCBR & 4e-22 & 0.371    & 1e-11 & 1e-27 \\ \hline
LIRA & 2e-4  & 4e-20 & 8e-21 & 2e-8  \\ \hline
\end{tabular}}
\end{center}
\end{table}


\begin{table}[]
\begin{center}
\caption{P-values from the Wilcoxon rank-sum test performed on the Dice score obtained by the MPUNet2+ versus other compared methods (viz., MPUNet, MPUNet3+, MPUNet2+ DS, and MPUNet3+ DS at 6-planes)  for OAI, CCBR, and LIRA datasets.} 
\label{wtest} 
\resizebox{\columnwidth}{!}{
\begin{tabular}{|l|c|c|c|c|}
\hline
\textbf{Dataset} &
  \textbf{\begin{tabular}[c]{@{}l@{}}MPUNet2+ \\ vs.\\ MPUNet\end{tabular}} &
  \textbf{\begin{tabular}[c]{@{}l@{}}MPUNet2+ \\ vs.\\ MPUNet3+\end{tabular}} &
  \textbf{\begin{tabular}[c]{@{}l@{}}MPUNet2+ \\ vs.\\ MPUNet2+ DS\end{tabular}} &
  \textbf{\begin{tabular}[c]{@{}l@{}}MPUNet2+ \\ vs.\\ MPUNet3+ DS\end{tabular}} \\ \hline
OAI  & 0.028    & 0.005    & 4e-30 & 4e-34 \\ \hline
CCBR & 2e-25 & 0.923    & 4e-12 & 1e-25 \\ \hline
LIRA & 4e-4  & 2e-29 & 4e-29 & 1e-17  \\ \hline
\end{tabular}}
\end{center}
\end{table}

\subsubsection{Qualitative evaluation}
\begin{figure}[ht]
    \resizebox{\columnwidth}{!}{
	\begin{tabular}{cccccc}
		\begin{subfigure}{0.151\textwidth}\centering\includegraphics[height=3cm, width=2cm]{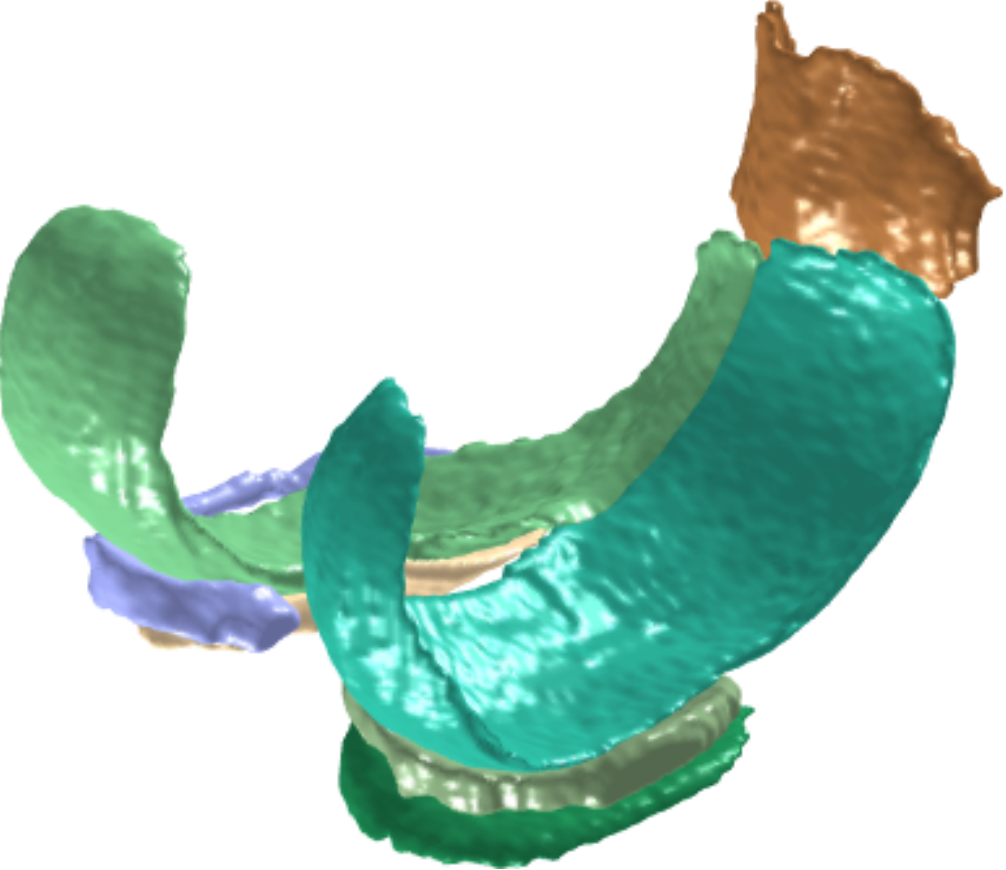}\end{subfigure}& 
		\begin{subfigure}{0.151\textwidth}\centering\includegraphics[height=3cm, width=2cm]{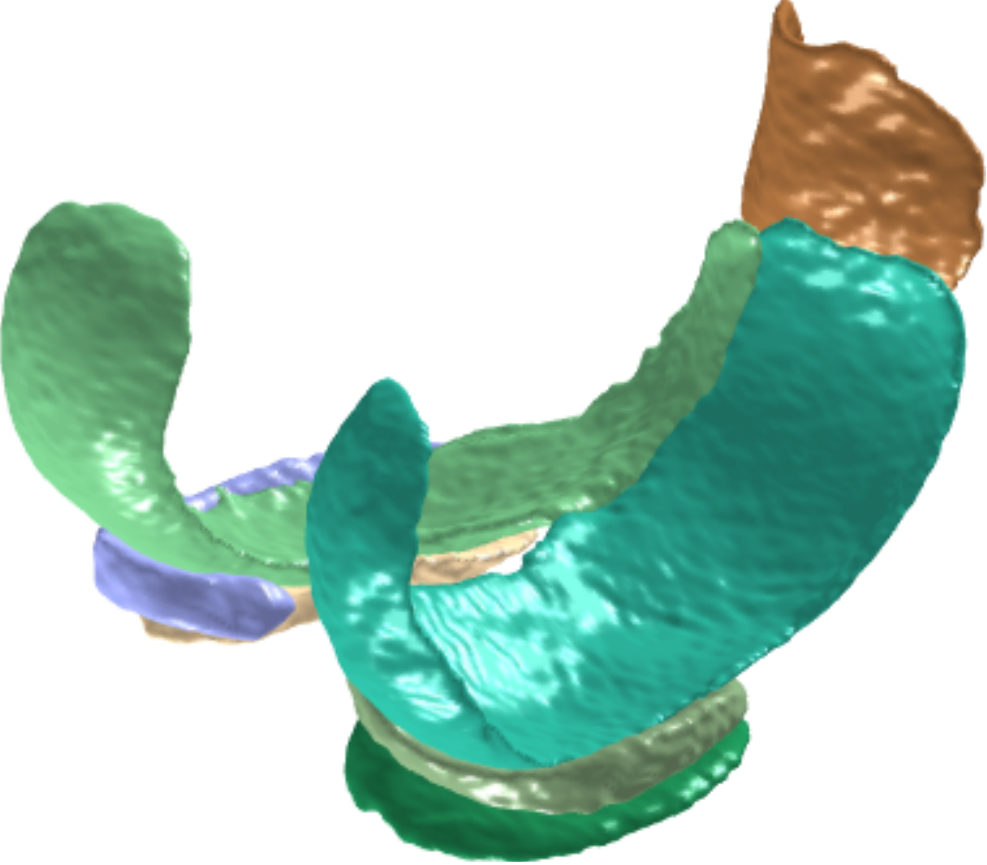}\end{subfigure}&
		\begin{subfigure}{0.151\textwidth}\centering\includegraphics[height=3cm, width=2cm]{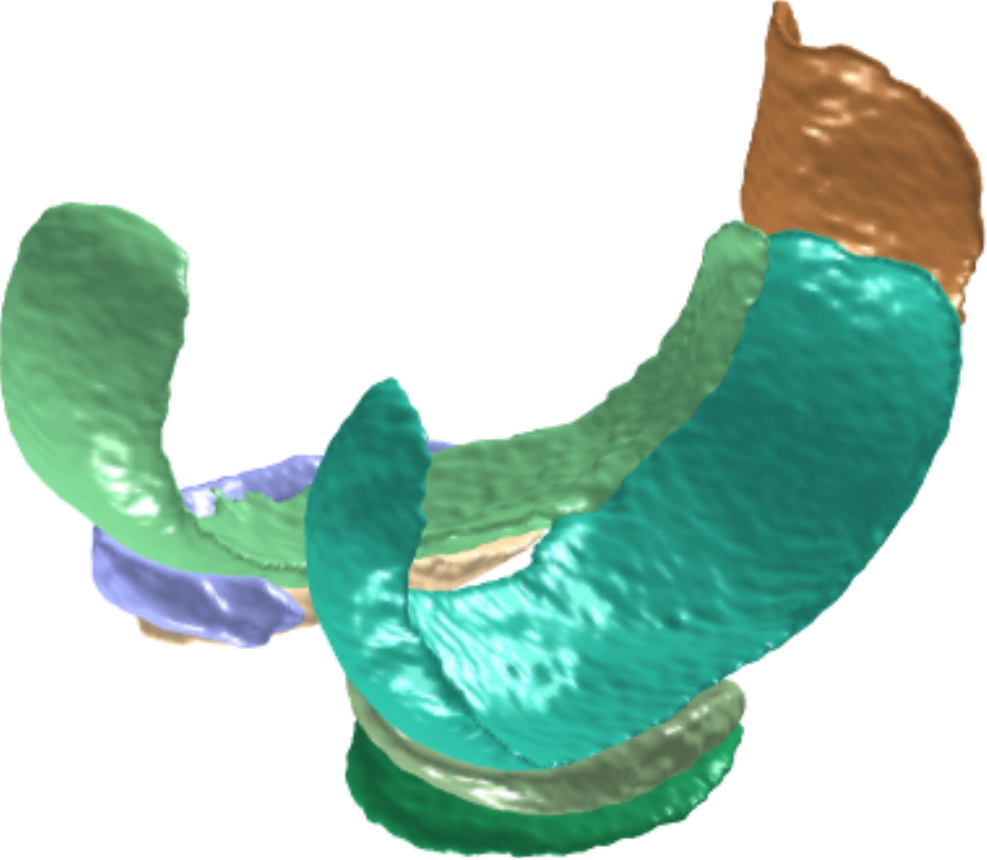}\end{subfigure}&
		\begin{subfigure}{0.151\textwidth}\centering\includegraphics[height=3cm, width=2cm]{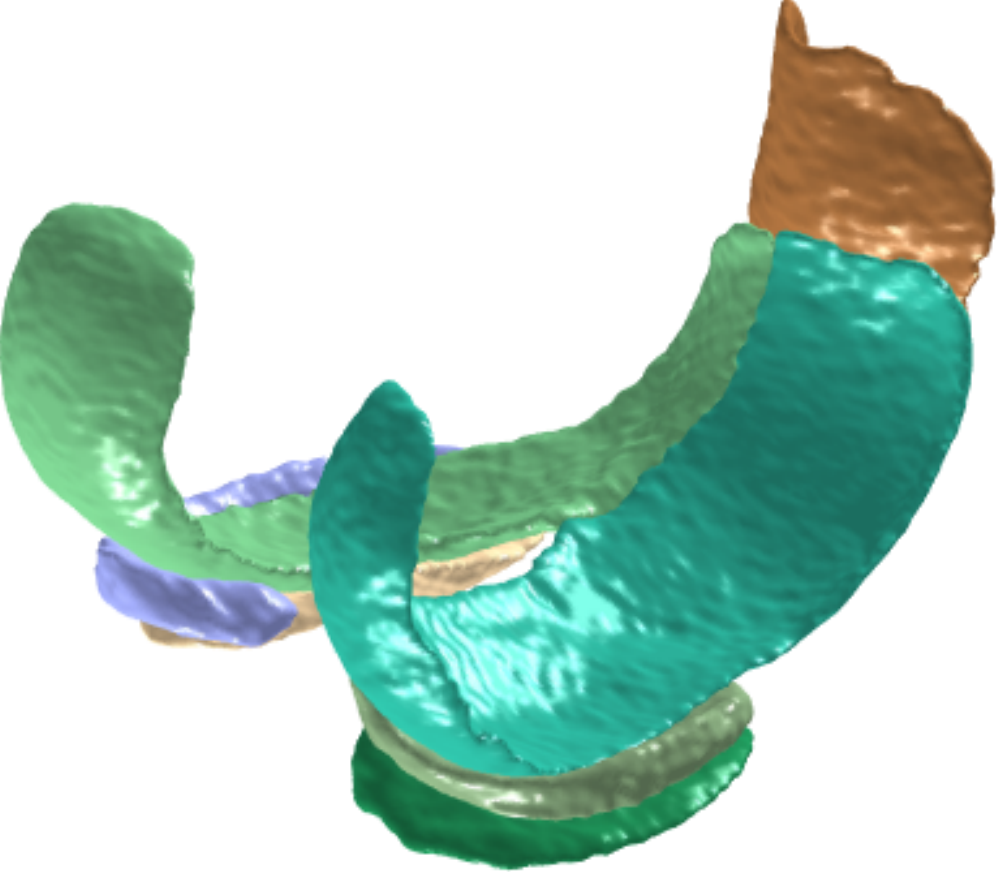}\end{subfigure}&
		\begin{subfigure}{0.151\textwidth}\centering\includegraphics[height=3cm, width=2cm]{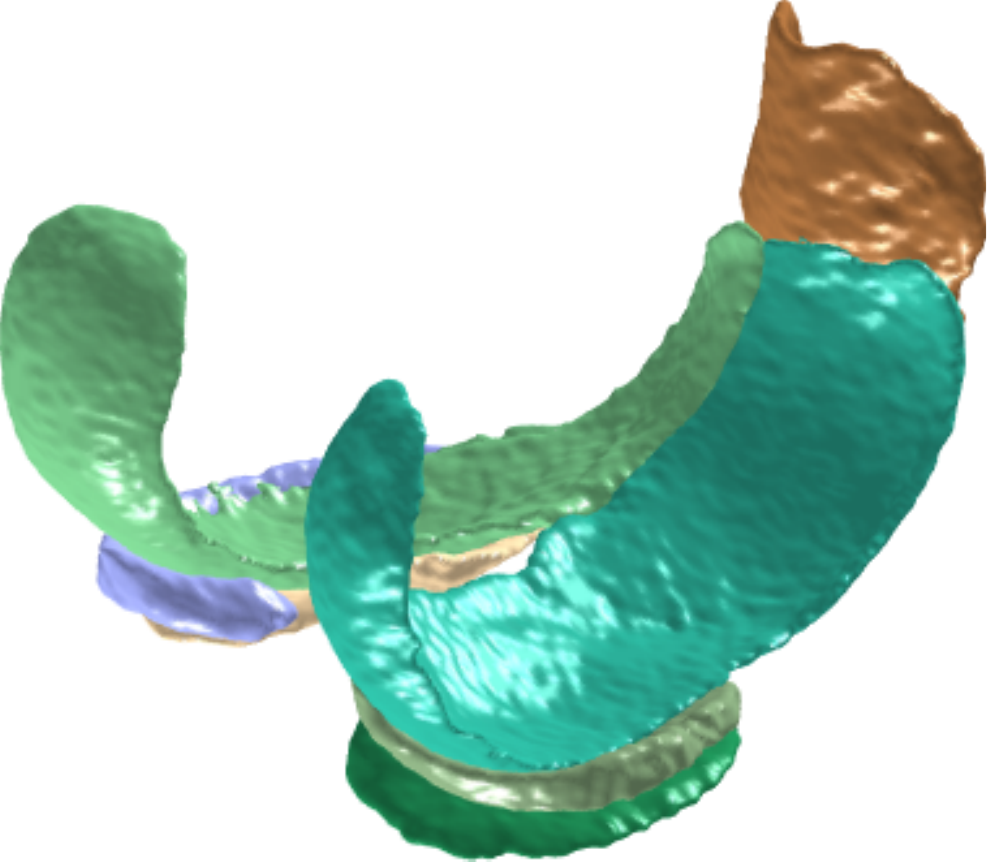}\end{subfigure}&
		\begin{subfigure}{0.151\textwidth}\centering\includegraphics[height=3cm, width=2cm]{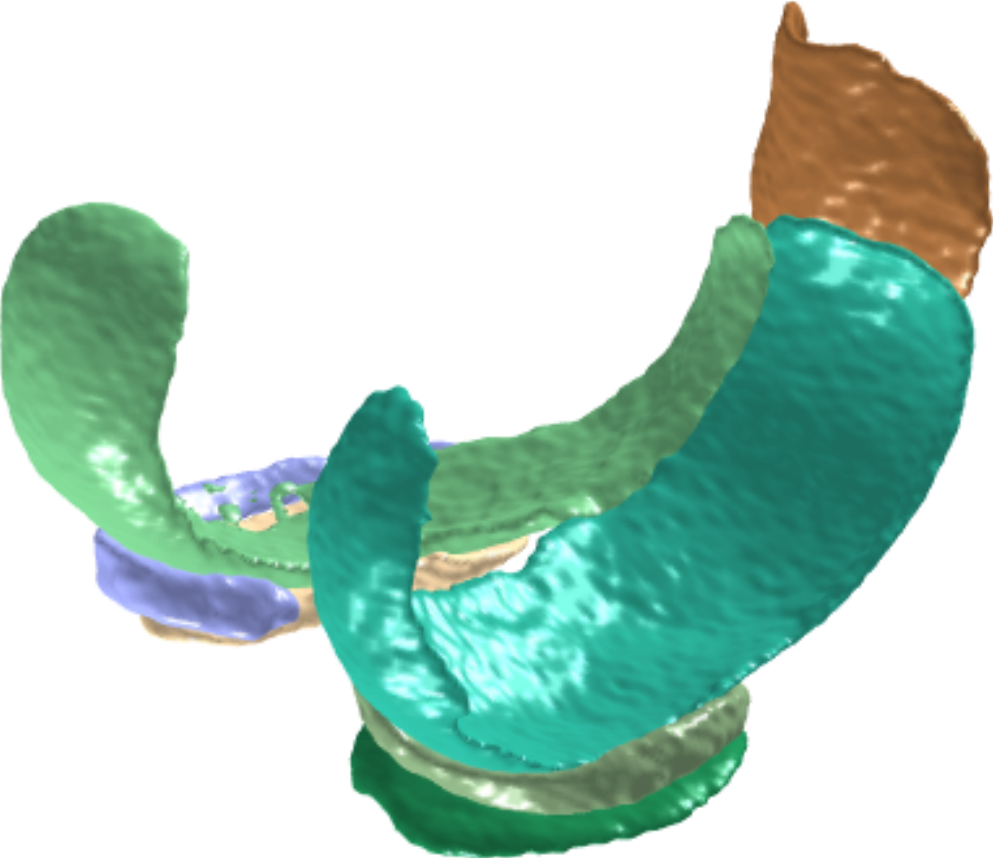}\end{subfigure}\\
		\begin{subfigure}{0.151\textwidth}\centering\includegraphics[height=3cm, width=2cm]{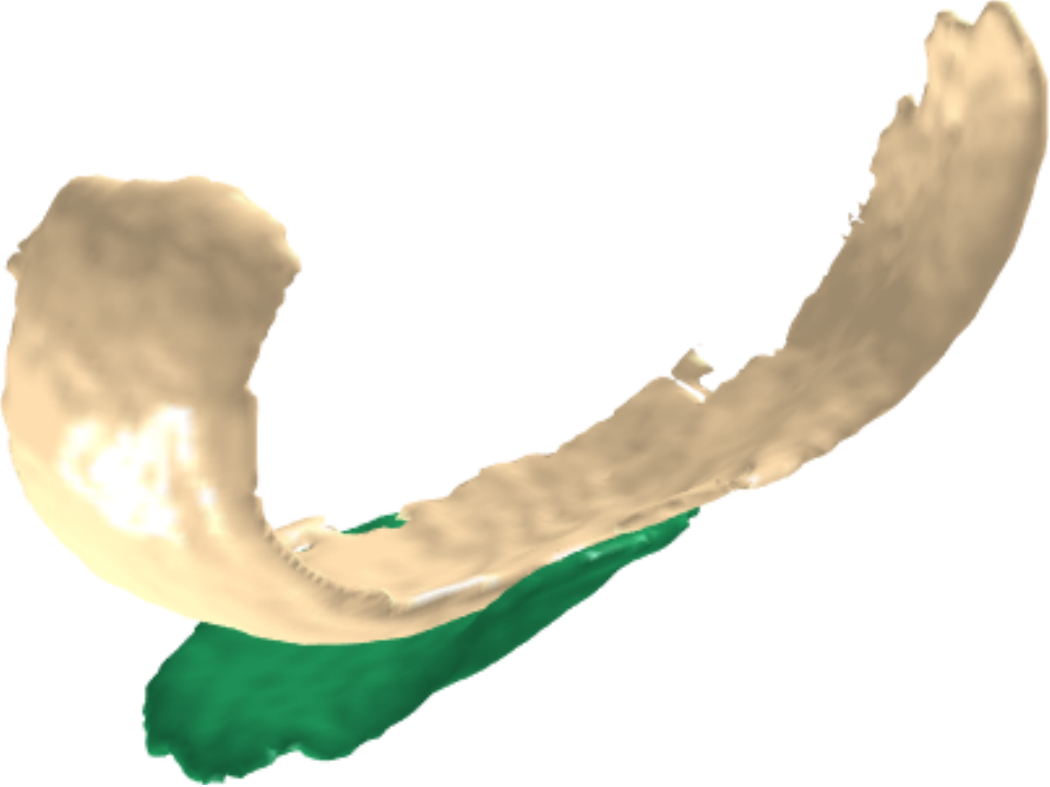}\end{subfigure}&
		\begin{subfigure}{0.151\textwidth}\centering\includegraphics[height=3cm, width=2cm]{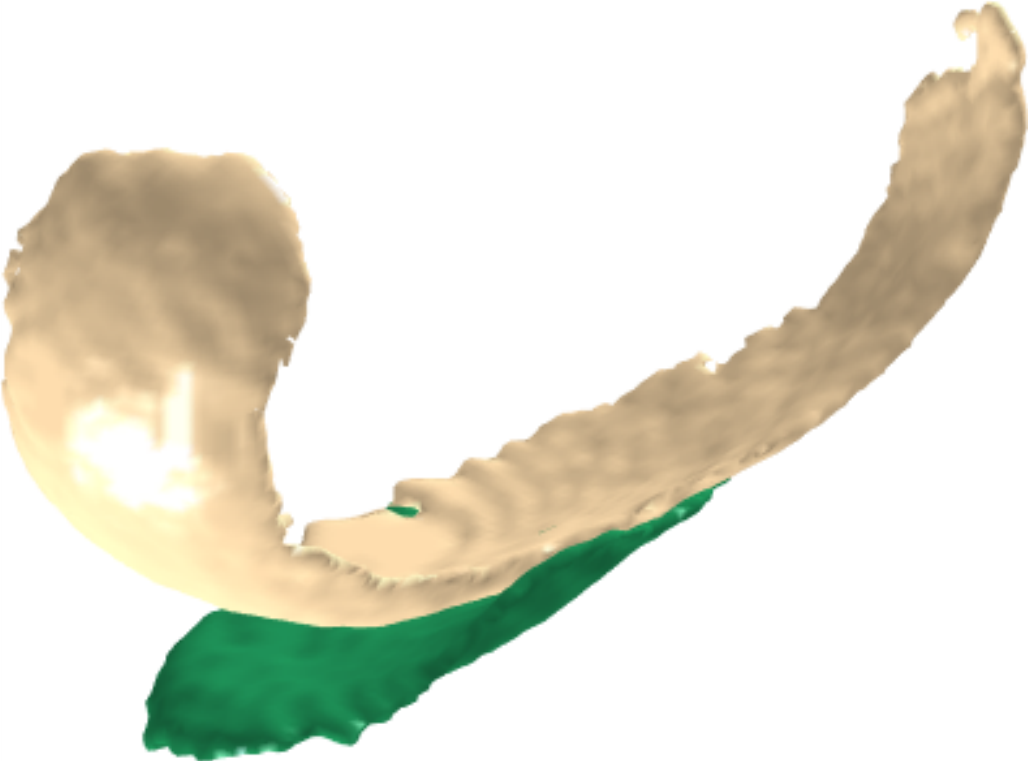}\end{subfigure}&
		\begin{subfigure}{0.151\textwidth}\centering\includegraphics[height=3cm, width=2cm]{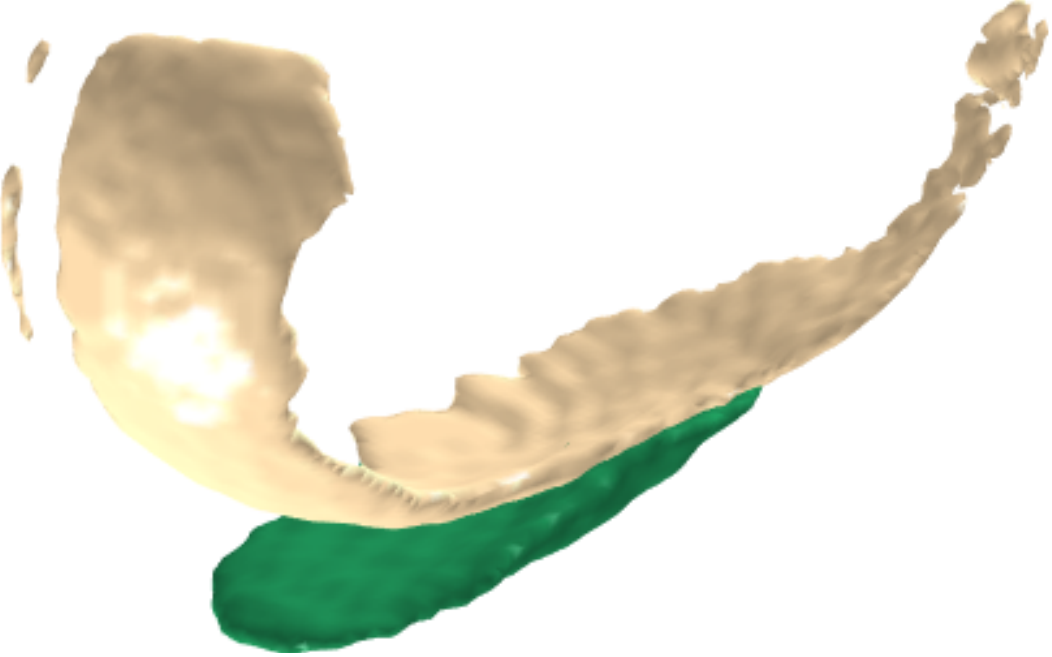}\end{subfigure}&
		\begin{subfigure}{0.151\textwidth}\centering\includegraphics[height=3cm, width=2cm]{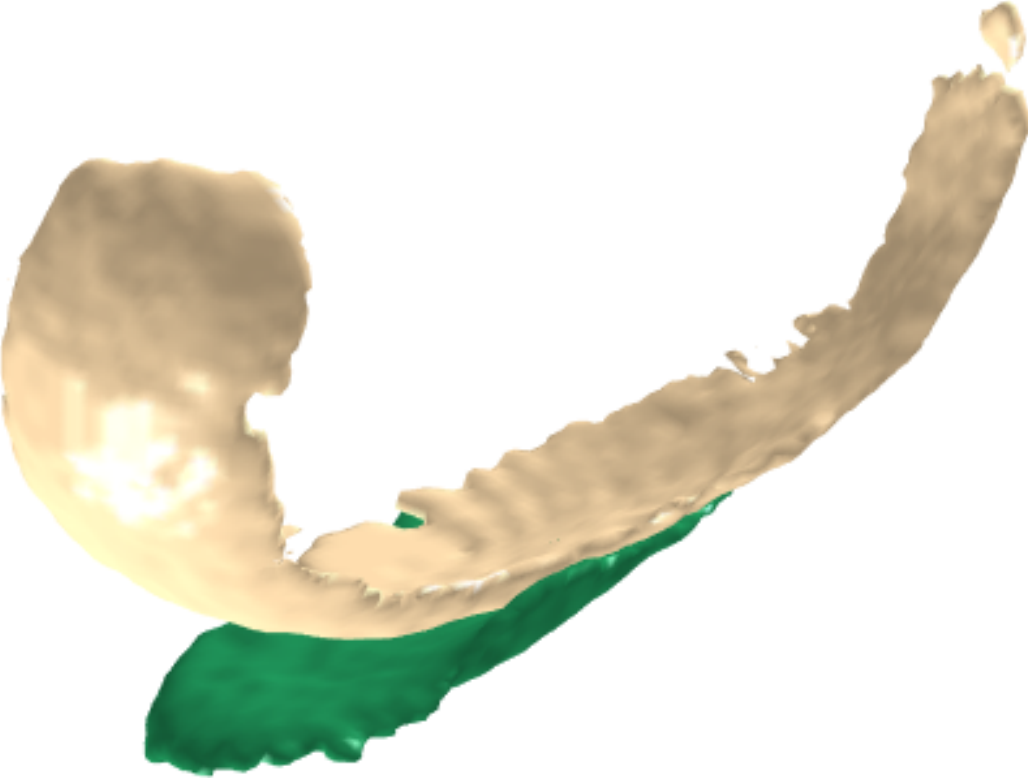}\end{subfigure}&
		\begin{subfigure}{0.151\textwidth}\centering\includegraphics[height=3cm, width=2cm]{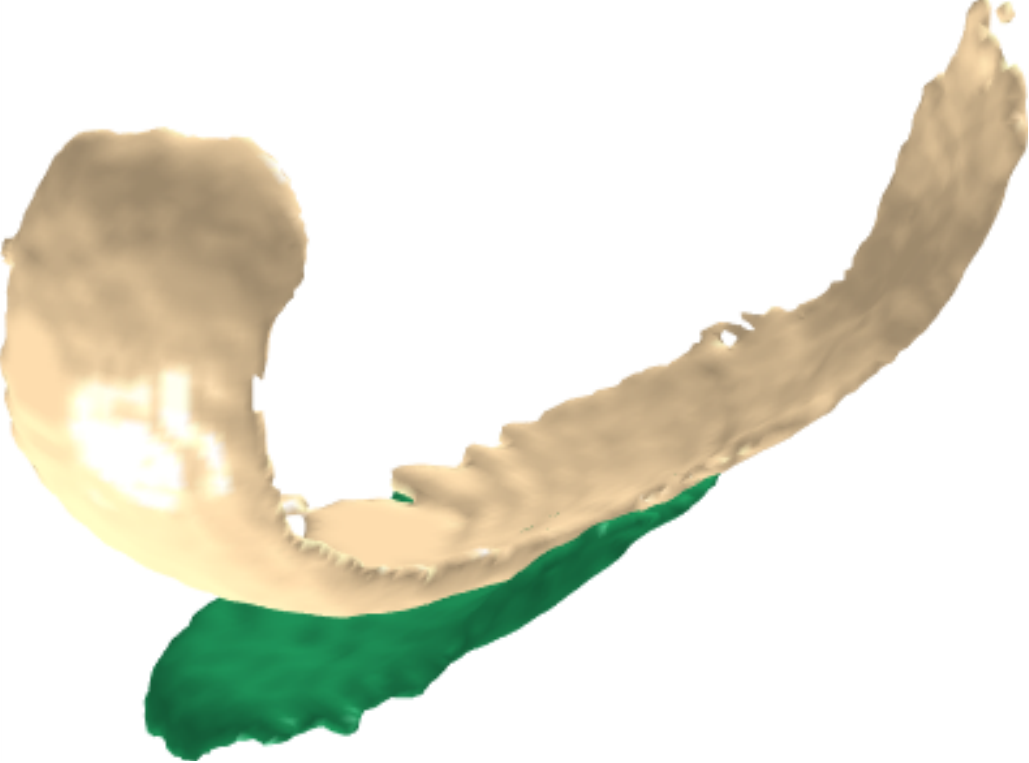}\end{subfigure}&
		\begin{subfigure}{0.151\textwidth}\centering\includegraphics[height=3cm, width=2cm]{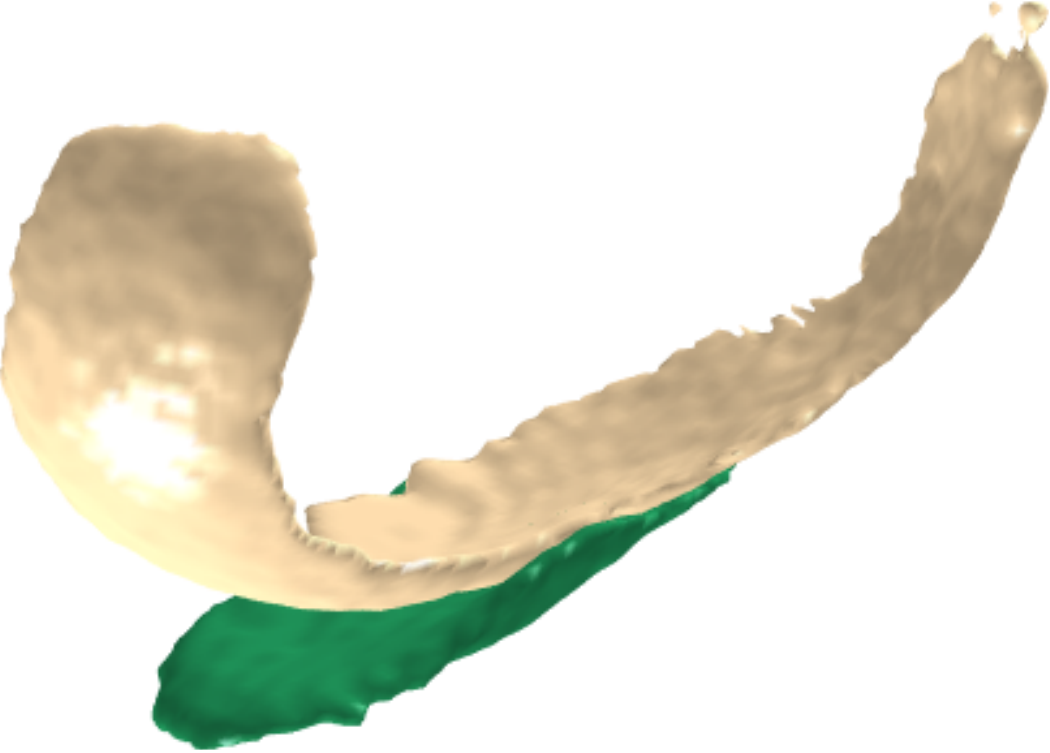}\end{subfigure}\\
		\begin{subfigure}{0.151\textwidth}\centering\includegraphics[height=3cm, width=2cm]{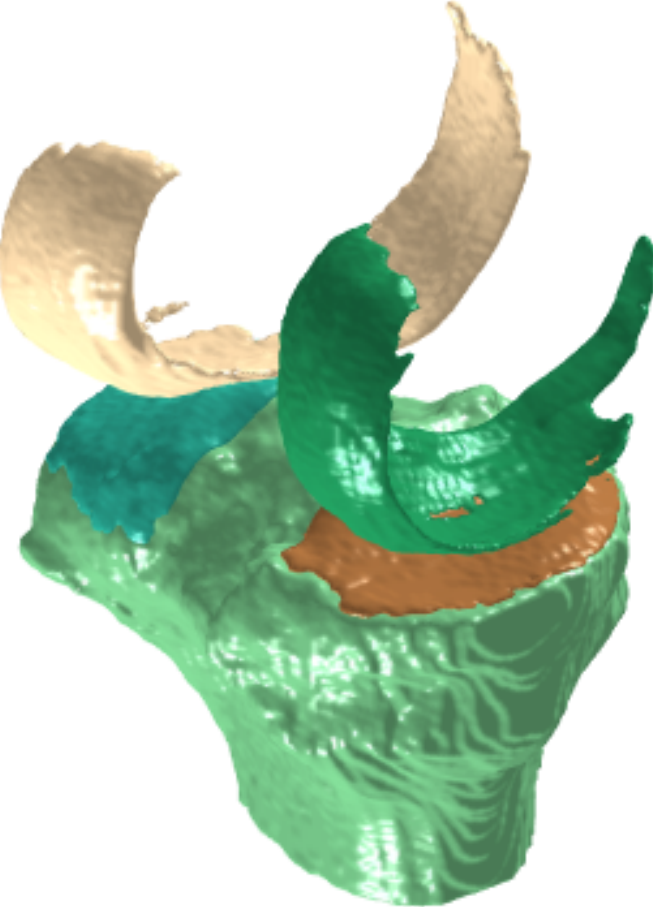}\caption{}\end{subfigure}&
		\begin{subfigure}{0.151\textwidth}\centering\includegraphics[height=3cm, width=2cm]{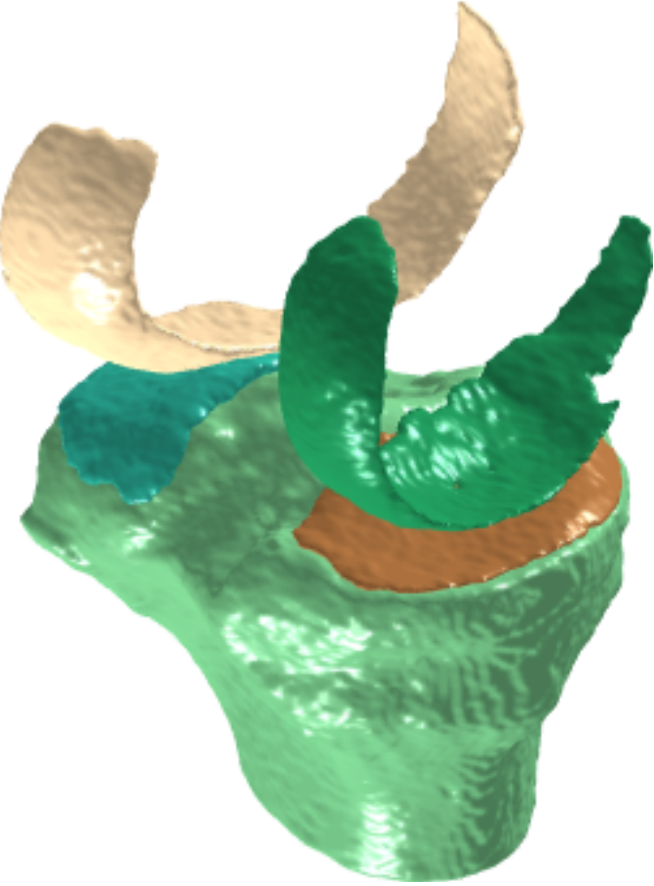}\caption{}\end{subfigure}&
		\begin{subfigure}{0.151\textwidth}\centering\includegraphics[height=3cm, width=2cm]{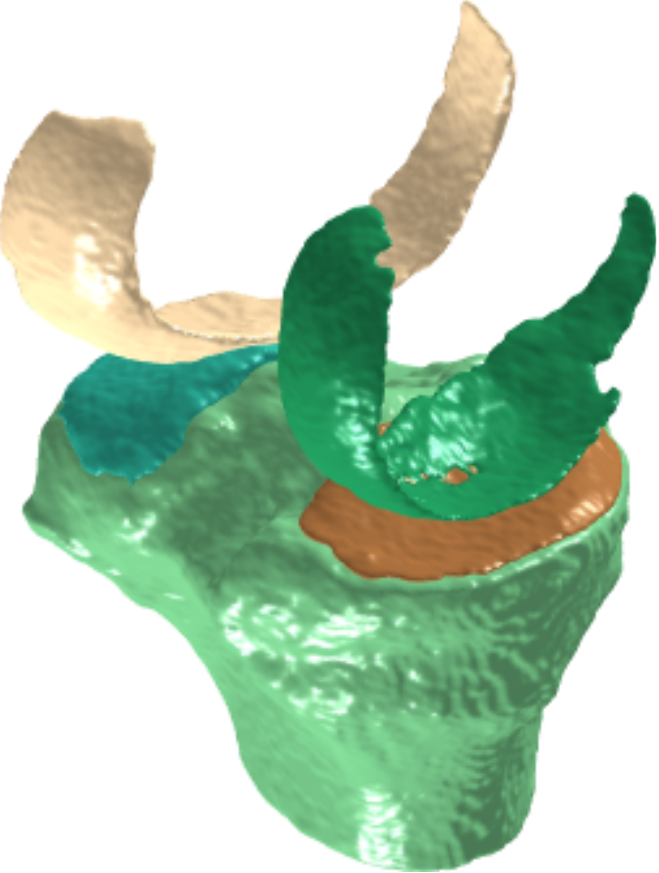}\caption{}\end{subfigure}&
		\begin{subfigure}{0.151\textwidth}\centering\includegraphics[height=3cm, width=2cm]{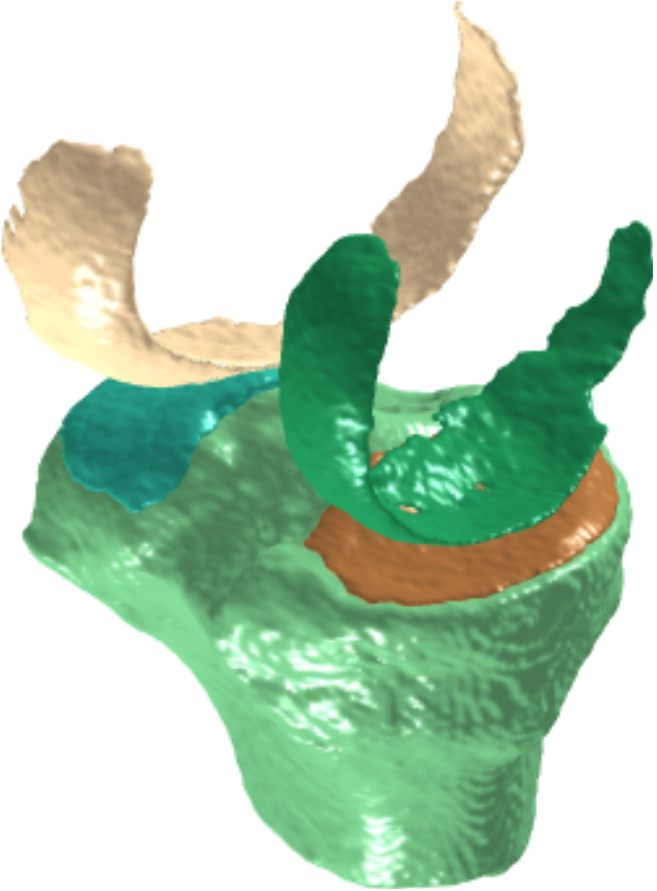}\caption{}\end{subfigure}&\begin{subfigure}{0.151\textwidth}\centering\includegraphics[height=3cm, width=2cm]{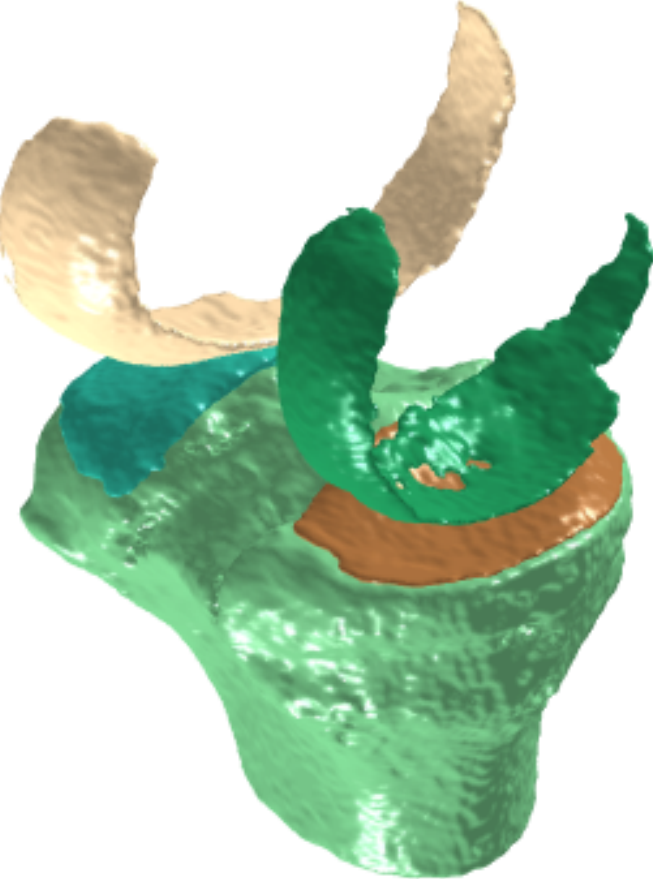}\caption{}\end{subfigure}&
		\begin{subfigure}{0.151\textwidth}\centering\includegraphics[height=3cm, width=2cm]{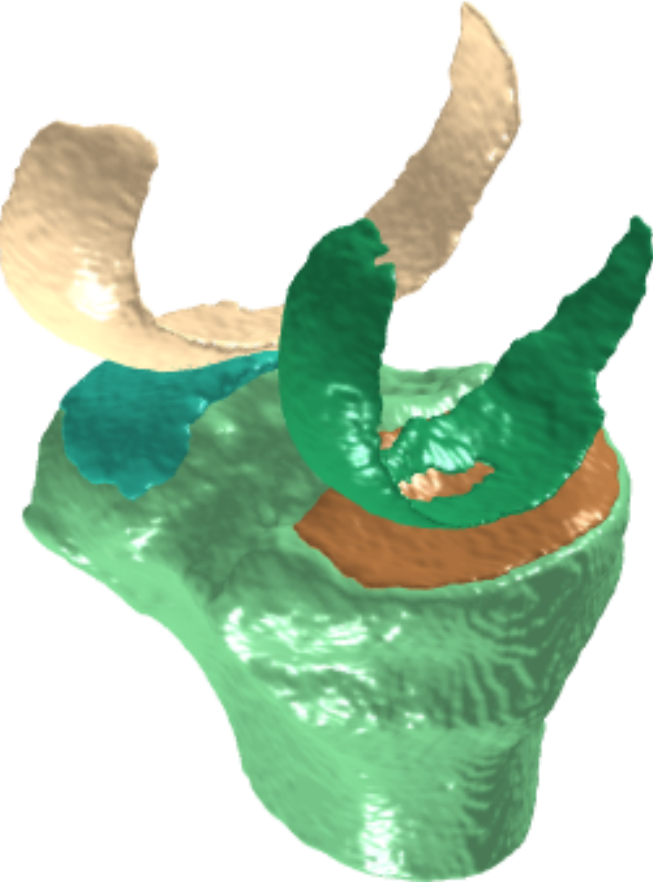}\caption{}\end{subfigure}
		
	\end{tabular}}
	\caption{ \label{visualization}
		Visual comparison of the performance in the form of full segmentation of all the methods obtained on a 2D slice corresponding to median dice score of all the datasets: where (a) Ground truth, (b) MPUNet2+, (c) MPUNet, (d) MPUNet3+, (e) MPUNet2+ DS, (f) MPUNet3+ DS; row wise, top to bottom row: OAI, CCBR, LIRA datasets.} 
\end{figure}

The qualitative evaluation for the performance comparison are shown in Fig.~\ref{visualization}. For the fair visualization, we have selected the single 2D slice from each dataset corresponding to the median dice score. The full segmentation result of all the available cartilages are displayed for the selected 2D slice. The columns of Fig.~\ref{visualization} show the ground truth and the results of MPUNet2+, MPUNet, MPUNet3+, MPUNet2+ DS, and MPUNet3+ DS respectively. The rows of the aforementioned figure show the results corresponding to OAI, CCBR, and LIRA datasets. It has been shown that for all the datasets, the visualization of MPUNet2+ is closer to ground truth as comparison to other compared methods i.e., all the cartilages in MPUNet2+ method are segmented well with minor error. 

Additionally, we have also done the cartilage-wise visual comparison of total count of false positives (FPs) and false negatives (FNs) from 3D segmented results of each cohorts for the MPUNet and MPUNet2+. For display purpose, we have projected the FNs and FPs counts from three dimension segmentation to two dimension. These results corresponding to (1) Medial Tibial, Lateral Tibial, Medial Femoral, Lateral Femoral, and Medial Meniscus for OAI (2) Medial Femoral and Medial Tibial for CCBR (3) Medial Tibial, Lateral Tibial, Medial Femoral, Lateral Femoral, and Tibia for LIRA dataset are shown in Figs. S1-S12 respectively. In the aforementioned figures, FNs and FPs are shown in first and second rows; MPUNet and MPUNet2+ results are displayed in first and second columns respectively. It is evident from above figures that, MPUNet2+ has significantly less number of FNs and FPs than MPUNet for each cartilages and cohorts.   
\begin{table}[]
\begin{center}
\caption{Statistical description of performance from box and whisker plot~\ref{box}.} \label{Statistical_box} 
\resizebox{\columnwidth}{!}{
\begin{tabular}{|l|c|c|c|c|c|c|c|}
\hline
\textbf{\begin{tabular}[c]{@{}l@{}}Cartilage/\\ Method\end{tabular}} &
  \textbf{Min} &
  \textbf{Max} &
  \textbf{Mean} &
  \textbf{\begin{tabular}[c]{@{}c@{}}25th \\ Percentile\end{tabular}} &
  \textbf{Median} &
  \textbf{\begin{tabular}[c]{@{}c@{}}75th \\ Percentile\end{tabular}} &
  \textbf{IQR} \\ \hline
\begin{tabular}[c]{@{}l@{}}Medial tibial\\ MPUNet2+\end{tabular}  & 0.622 & 0.918 & 0.857 & 0.837 & 0.866 & 0.885 & 0.048 \\ \hline
\begin{tabular}[c]{@{}l@{}}Medial femoral\\ MPUNet2+\end{tabular} & 0.589 & 0.922 & 0.836 & 0.810 & 0.848 & 0.875 & 0.065 \\ \hline
\begin{tabular}[c]{@{}l@{}}Medial tibial\\ MPUNet\end{tabular}    & 0.543 & 0.916 & 0.843 & 0.824 & 0.852 & 0.878 & 0.054 \\ \hline
\begin{tabular}[c]{@{}l@{}}Medial femoral\\ MPUNet\end{tabular}   & 0.524 & 0.918 & 0.819 & 0.787 & 0.827 & 0.866 & 0.079 \\ \hline
\end{tabular}}
\end{center}
\end{table}

\begin{table}[ht]
\begin{center}
\caption{Compartment-wise performance (Dice-score) of MPUNet2+ 6 planes over OAI, CCBR, and LIRA datasets. "---" represents the absence of manual segmentations for this compartment in the dataset} \label{Cartilage}
\begin{tabular}{|l|c|c|c|}
\hline
\textbf{Cartilage} & \textbf{OAI} & \textbf{CCBR} & \textbf{LIRA} \\ \hline
Medial Tibial      & 0.862        & 0.849         & 0.876         \\ \hline
Lateral Tibial     & 0.905        & ---           & 0.896         \\ \hline
Medial Femoral     & 0.875        & 0.835         & 0.782         \\ \hline
Lateral Femoral    & 0.893        &---              & 0.801         \\ \hline
Patellar           & 0.815        &---               & ---              \\ \hline
Lateral Meniscus   & 0.888        &---               &---               \\ \hline
Medial Meniscus    & 0.836        & ---              & ---              \\ \hline
Tibia              &---              & ---              & 0.983         \\ \hline
\end{tabular}
\end{center}
\end{table}

\section{Discussion}\label{sec:discusion}

With the help of multiplanar model~\cite{perslev2019one}, we have evaluated four different UNet variants for segmentation of the cartilages from 3D knee MRI images. To better generalize the performance, these architectures have been experimented with 1, 3 and 6 number of planes and validated over 3 different datasets (OAI, CCBR, and LIRA). Quantitative evaluations can be summarised as: (1)  Table~\ref{performance_all} shows the dice score, computation time, and number of parameters in each methods. It is evident from above table (a) the dice score of MPUNet2+ is significantly better or near to significantly better in comparison to other methods for all datasets. Furthermore, on average 6 planes case of MPUNet2+ is better than 1, 3 Planes. (b) The computation time of MPUNet2+ is typically less than for the other methods. (c) There are fewer parameters (and thereby the less GPU memory requirement) in MPUNet2+ than MPUNet. However, it is more than MPUNet3+ and MPUNet3+DS. (2) To test the significant better performance of MPUNet2+ 6 planes, the results of paired t-test and Wilcoxon paired rank-sum test are displayed in Tables~\ref{ttest} and~\ref{wtest} respectively. For all cases except with MPUNet on OAI (for paired t-test), MPUNet3+ with CCBR (for paired t-test), and MPUNet3+ with CCBR (for Wilcoxon paired rank-sum test) the p-values are less than .05 i.e. MPUNet2+ performs significantly better. (3) For better generalization, box and whisker plot has been displayed in Fig.~\ref{box}. This plot has been calculated for 6 planes version of MPUNet and MPUNet2+ with Medial tibial and  Medial femoral cartilages of all three datasets combined together. The statistical description of above plot has been shown in Table~\ref{Statistical_box}. It is evident from aforementioned table that minimum, maximum, mean, 25th percentile, median, 75th percentile are higher in case of MPUNet2+. (4) Next, individual cartilage performance is also better for MPUNet2+ 6 planes (see Table~\ref{Cartilage}). (5) Furthermore, Liu et al.~\cite{liu2018deep} proposed a 2D slices based Knee cartilages segmentation approach from MRI. However, the performance of this method is not better than our MPUNet2+. (6) As per the literature, Dam et al.~\cite{dam2017simple} presented a method for automatic segmentation of high- and low-field knee MRI. This method placed second for cartilage segmentation in the SKI10 challenge. However, our method performs better than this for most of the cartilages.

The visualizations offer qualitative evaluations that can be summarised: (1) Fig.~\ref{visualization} displays the segmented results of selected 2D slice (for better comparison, this slice has been selected corresponding the median dice score) of OAI, CCBR and LIRA datasets for all the tested approaches. Again, it is shown from figure, the results of MPUNet2+ are more similar to the ground truth for all datasets.  (2) Furthermore, individual cartilage-wise total number of false negatives (FNs) and false positives (FPs) are also projected from complete 3D segmented cohort to 2D plane. The results have been displayed in supplementary section and show that MPUNet2+ has both fewer FNs and FPs than MPUNet.   

Deep Learning methods have been accused of being sensitive to domain shifts. When adapting a sequence for a new setup, there will be a small domain shift between the sites and the specific scanner versions (e.g. differing in software version). Obviously, the domain shift is much larger when changing sequence, scanner vendor, and field strength. Thereby, our validation with OAI and LIRA datasets has the ability to reveal whether small domain shifts would be manageable whereas very large domain shifts would be problematic (despite of their same scanning sequence). 

From the above experiments with extensive quantitative and qualitative evaluations, we observed that the MPUNet2+ significantly improves the results on every dataset. 
Our method can work efficiently with small amount of dataset and without the need of high computational power. We have fixed set of hyperparameter set i.e. model doesn't requires task specific information. We have extracted the 2D slices from different orientations i.e. most of the important information is available for training and testing steps. The project is available open source and does not need deep learning expertise to use i.e. good for non-technical users. 
Although the performance of MPUNet2+ is significantly better than MPUNet and some of the UNet3D models. However, MPUNet2+ doesn't perform better than few of the variants of UNet3D models. But, these 3D models require high computational power.

\section{Conclusions}\label{sec:Conclusions}
We have verified four different UNet inspired architectures using multiplanar UNet to produce a generalized method for accurate 3D knee cartilage segmentation. The proposed work has been applied on three different knee MRI datasets (OAI, CCBR, and LIRA). The quantitative and qualitative evaluation demonstrated that MPUNet2+ had higher accuracy, considerably fewer parameters, and less training time than other approaches for all the tested datasets. In the multi-planar 3D framework, the results demonstrated that MPUNet2+ is likely to outperform MPUNet. However, since this is very different from the setting where the UNet2+ was originally proposed, these results support that the Unet2+ should in general be considered as an alternative to the golden standard Unet. 

\section*{Declaration of competing interest}\label{sec:Conflict}
The authors have no conflicts of interest to declare. 

\section*{Acknowledgement}
In this study, the work on the LIRA cohort was supported by an unrestricted grant from Novo Nordisk A/S including active and placebo medicine, and by the Cambridge Weight Plan donating dietary supplements. In addition, the work on the LIRA cohort was supported by a core grant from the Oak Foundation (OCAY-13-309) given to the Parker Institute, Copenhagen University Hospital, Bispebjerg and Frederiksberg. The funders had no influence on the design or conduct of the study and were not involved in data collection or analysis, in the writing of the manuscript, or in the decision to submit it for publication.

\end{document}